\documentclass[preprint]{revtex4}

\usepackage{graphicx}

\newcommand{\beq}{\begin{equation}}
\newcommand{\enq}{\end{equation}}
\newcommand{\bea}{\begin{eqnarray}}
\newcommand{\ena}{\end{eqnarray}}
\newcommand{\ad}{a^{\dag}}

\begin{document}

\title{Dipole and monopole modes in the Bose-Hubbard model in a trap}%
\author{Emil Lundh}
\affiliation{Department of Physics, Royal Institute of Technology,
AlbaNova, SE-106 91 Stockholm, Sweden}

\begin{abstract}
The lowest-lying collective modes of a trapped Bose gas in an
optical lattice are studied in the Bose-Hubbard model. An exact
diagonalization of the Hamiltonian is performed in a
one-dimensional five-particle
system in order to find the lowest few eigenstates.
Dipole and breathing character of the eigenstates is confirmed in
the limit where the tunneling dominates the dynamics, but under
Mott-like conditions the excitations do not correspond to
oscillatory modes.
\end{abstract}
\maketitle
\section{Introduction}

Exciting collective modes is a useful and popular tool for 
probing the
many-body physics of trapped atomic gases. Following the first
creation of a trapped condensate in 1995 \cite{anderson1995a},
modes in these systems have been
subject to extensive theoretical and experimental study
\cite{pethick2001a}. The fundamental zero-temperature
theory was
laid down about half a century ago \cite{bogoliubov1947a},
and was readily adapted to the case of trapped condensates
\cite{stringari1996a}. In three dimensions, oscillatory modes are
naturally classified according to their multipolarity, and can be
selectively excited by deforming the magnetic trap, or
applying laser pulses that repel or attract
the atoms in selected regions of space \cite{jin1996a,mewes1996a}.

The picture is complicated considerably if one adds an
optical lattice, consisting of one or several standing laser
waves that act as a spatially periodic potential on the atoms.
In the limit of a weak optical potential, the mode frequencies
are simply given by those of the trapped Bose-Einstein condensed
cloud in the absence of an optical lattice, but
renormalized by the effective mass acquired by the bosons in the
periodic potential
\cite{cataliotti,kraemer2002a}.
Away from this limit, however,
the presence of an optical lattice offers quite different physics,
and a new phase appears, namely the Mott insulator, when the 
interactions are strong \cite{greiner2001a,dries}.
In addition, when an external trapping potential is present,
there exist parameter
regimes where spatially separated regions
of Mott-insulating and superfluid behavior coexist
\cite{jaksch,batrouni2002a}.
The behavior of the trapped gas and the nature
of its collective modes are expected to
become quite different in these regimes compared to the quite
well understood case of a trapped cloud with no optical lattice
present \cite{pupillo2003a}.

In order to be able to address both the strongly and weakly
interacting case and the crossover between these, we shall study the
Bose-Hubbard model in the exactly solvable case of few bosons and in
one dimension. The method shall be exact diagonalization in a truncated
basis. This way we hope to gain qualitative knowledge of the
spectrum that applies also to larger systems and higher dimensions.
The paper is organized as follows.
The Hamiltonian and the numerical method are explained
in Sec. \ref{sec:model}. The nature of the ground state and the
low-lying excitations in a
shallow trap is investigated in Sec.\ \ref{sec:shallow} and
the case of a tight trap in Sec.\ \ref{sec:tight}. Concluding
remarks are given in Sec.\ \ref{sec:conclusions}.

\section{Bose-Hubbard model and truncated basis}
\label{sec:model}

The starting point is the Bose-Hubbard Hamiltonian \cite{jaksch}
\beq\label{hamiltonian}
H = \sum_{r} \frac{U}{2} \ad_r\ad_r a_r a_r -
\frac{J}{2} \ad_r (a_{r+1}+a_{r-1})
+\frac{\omega^2 r^2}{2} \ad_r a_r.
\enq
The first term in this Hamiltonian describes the interactions
which are effectively repulsive if $U>0$ (which is the case in
this paper),
the second, so-called tunneling or hopping term is associated with the
kinetic energy, and the last term describes the external trapping
potential. The index $r$ denotes the spatial position and takes on
integer values.
Such a one-dimensional Hubbard model describes
a system with a tight trap in the directions perpendicular
to the lattice so that the other degrees of freedom are frozen
out, thus resembling a coupled chain of quantum dots. Higher
dimensions will make the picture more complicated, but
the main qualitative features observed in the present paper are 
expected to carry over to higher dimensions.

The Hamiltonian contains three physical parameters. The tunneling
strength $J$ can be written
\beq
\label{jdef}
J = \frac{\hbar^2}{m^* \delta r^2},
\enq
where $\delta r$ is the spacing between wells and $m^*$ is the effective
mass acquired by the atoms due to the periodic potential
\cite{kraemer2002a}.
The interaction strength $U$ is related to trap parameters through
the relation
\beq
U = \frac{4\pi\hbar^2a}{m}\int d^3r |\Psi_{\rm TB}(r)|^4,
\enq
where $a$ is the {\it s}-wave scattering length and $\Psi_{\rm TB}$
is the ground-state wave function in one potential well in the
tight-binding approximation. The effective trap frequency
$\omega$ is defined in terms of the bare particle mass $m$ and
trap frequency $\Omega$ as
\beq
\label{omegadef}
\omega = \sqrt{m}\Omega \delta r.
\enq
Let us at this point rescale the Hamiltonian and
work in units of $J$; formally we set $J=1$ and retain $U$ and
$\omega$ as the two parameters of the system.
In addition to the parameters already discussed, the number
of atoms $N$ or alternatively the chemical potential is a 
parameter of the system; we shall fix $N=5$ in this study.
Furthermore, for a few-particle system the even/odd
parity of the number of sites $L$ may also
play a decisive role; such effects vanish in the limit of large
systems. 
For definiteness only odd $L$ will be considered, but some 
attention will be paid to parity effects where appropriate.

In a number-conserving formalism, the natural basis is the
set of real-space Fock states that are also eigenstates 
of the interaction and trap energies:
\beq
|\ldots n_{r-1} n_r n_{r+1} \ldots\rangle = \ldots (\ad_{r-1})^{n_{r-1}}
(\ad_{r})^{n_{r}}(\ad_{r+1})^{n_{r+1}}\ldots |0\rangle.
\enq
The trapping potential implies a finite system size: it
turns out that between one and 25 sites is needed to accommodate
a system of five particles for the trap parameters considered here.
The size of the basis for a system of size $L$ with $N$
particles is $(N+L-1)!/[N!(L-1)!]$; for $N=5$, $L=25$ 
the number of
states is 118755. Clearly, the computations can be made much more
efficient if the basis is truncated so that the many improbable
Fock states do not contribute:  it is immediately obvious that
states such as $|N000\ldots 0\rangle$, where all the particles are
concentrated at one endpoint of the lattice, make only a very small
contribution to the dynamics.

There is, therefore, much to be gained if the basis is truncated.
The following scheme turns out to be practical for both strong and 
weak coupling, although it was designed for dealing with
Mott-insulator-like conditions where the interactions are strong.
Start with a single Fock state labeled $|1'\rangle$, for instance
the state with all
the particles at the same site, $|1'\rangle=|\ldots 00N00\ldots\rangle$.
Now enumerate all the states, $|2'\rangle,\ldots, |n'_1\rangle$, that can be
constructed from $|1'\rangle$ by one application of the tunneling
term in the Hamiltonian (\ref{hamiltonian}), and construct the 
Hamiltonian matrix elements in the
process. Operate again with the tunneling Hamiltonian on the
states $|2'\rangle,\ldots,|n'_1\rangle$ to form new states
$|n'_1+1\rangle,\ldots,|n'_2\rangle$, taking care not to
double-count states; iterate this step $p$ times so that a basis
is formed that consists of $n'_p$ states. Within this basis, the
Hamiltonian is now diagonalized and the ground state is found.
Among the $n'_p$ Fock states in the preliminary basis, choose the
one that has the largest overlap with the ground-state eigenvector
and label it $|1''\rangle$. Now discard all the other Fock states
that were just constructed, and instead
iterate the whole scheme again to
construct a new basis $|1''\rangle,\ldots,|n''_p\rangle$; do the
iteration a few (say, $M=3$ or 4) times. The basis thus
constructed, $|1^{(M)}\rangle,\ldots,|n^{(M)}_p\rangle$, will
contain all the Fock states that have significant overlap with the
ground state for the given physical parameters; in a sense, by
constructing this basis tunneling effects to $p$th order have been
incorporated. Convergence with
respect to $p$ and $M$ is readily checked, so that the
diagonalization can for all practical purposes be considered
exact.

The diagonalization is performed with ARPACK, which uses an
Arnoldi algorithm.

\section{Modes in shallow traps}
\label{sec:shallow}

\begin{figure}
\includegraphics[width=\columnwidth]{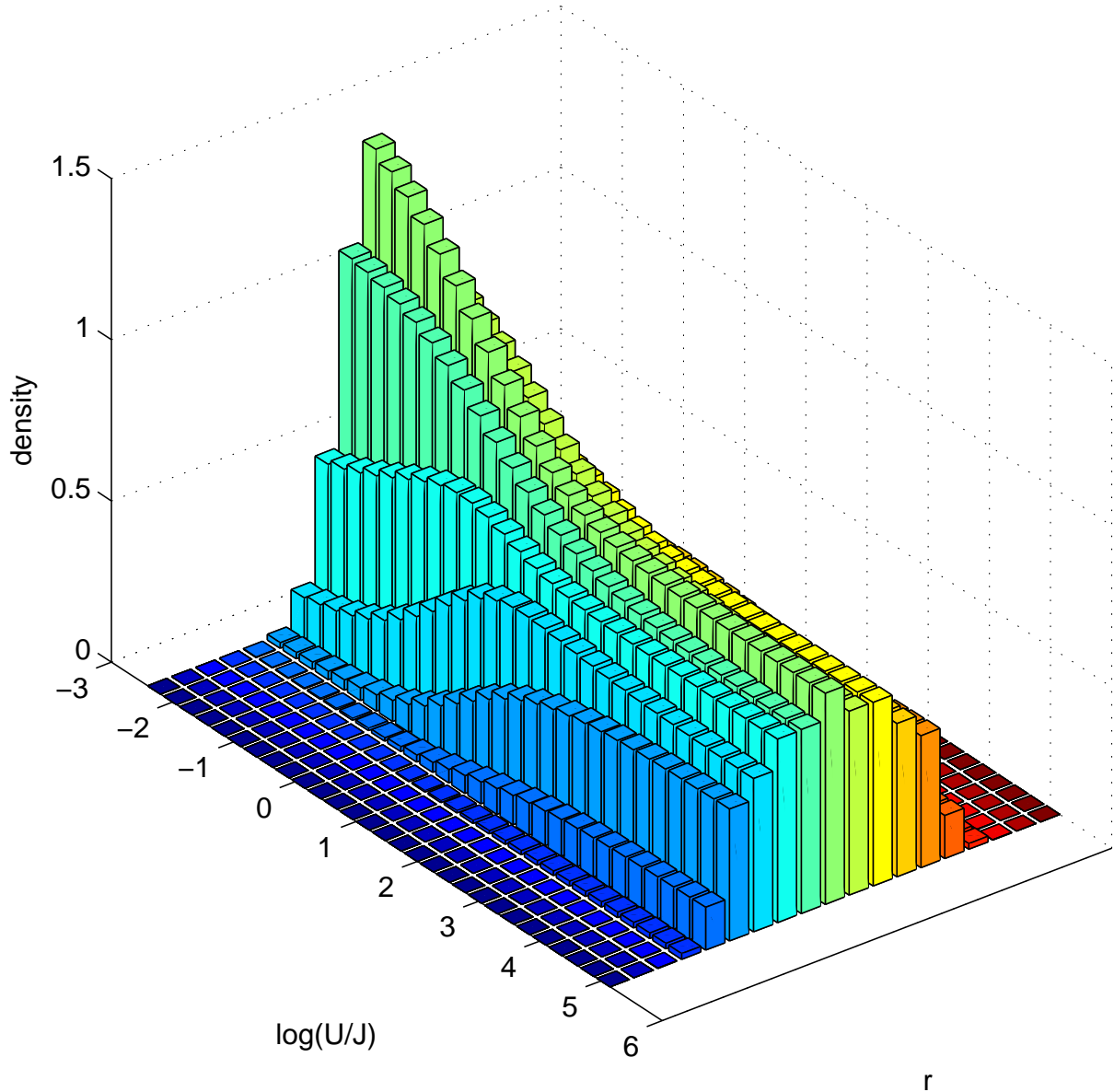}
\caption[]{Ground-state density profile for different values of
on-site interaction $U$ in a shallow trap with $\omega=0.3$.
\label{fig:gdensshallow}}
\end{figure}

\begin{figure}
\includegraphics[width=\columnwidth]{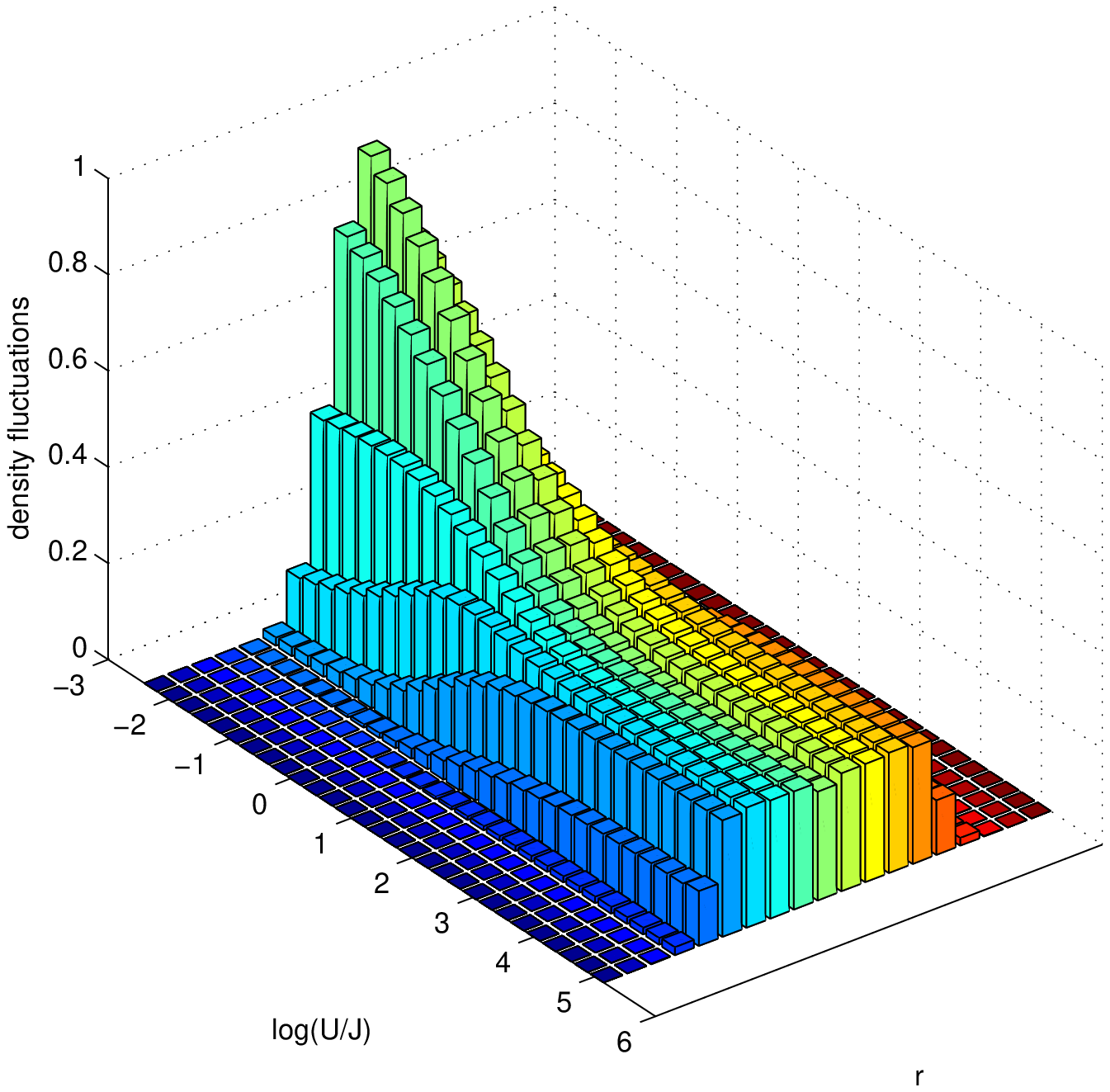}
\caption[]{Quantum fluctuations in the ground-state density,
$\langle\delta n\rangle=\langle n^2\rangle-\langle n \rangle^2$,
for different values of $U$ in the shallow trap, with
$\omega=0.3$.
\label{fig:fluctshallow}}
\end{figure}

The competition between the tunneling, interaction and trap
energies gives rise to a rich phase diagram (cf.\ \cite{jaksch,dries}).
Consider first the shallow trap. Figure \ref{fig:gdensshallow}
displays the ground-state density distribution for values of $U$
ranging from weak to strong interactions, with $\omega=0.3$.
The quantum fluctuations of the number of particles in the central
well are displayed in Fig.~\ref{fig:fluctshallow}. It is
seen that we are in the fluctuation-dominated, superfluid regime.
The very slight suppression of fluctuations in the centermost well
is in fact a signal that we are in the vicinity of the Mott 
insulating regime; 
if the trapping strength is increased, this suppression
becomes stronger, as shall be discussed in Sec.\ \ref{sec:tight}.
For the present trapping strength, $\omega=0.3$, the effect is
barely noticeable even for $U=100$.

In this shallow trap, the competition between the tunneling
and interaction is decisive in the interior of the system, while
the trap still determines the size and the behavior close to the
boundary. The density is spread out over many sites, and
therefore the system resembles a trapped, Bose-condensed cloud in
the absence of a lattice.
It is in this ``trapped BEC'' regime
that the experiments of Ref.\ \cite{cataliotti} were conducted.
While five particles are too few to be considered truly in
the trapped BEC regime,
the results reported here still give a hint of that limit, as is
seen in Fig.\ \ref{fig:gdensshallow}.
In a trapped condensate in the absence of an optical lattice,
the density distribution is Gaussian for
weak coupling, but flattens out and takes on the shape of an 
inverted parabola
for stronger coupling \cite{pethick2001a}. The density
profiles shown in Fig.\ \ref{fig:gdensshallow} can of course
not be expected to exactly follow this behavior, because of the
discreteness, but the
dependence on coupling is similar.

\begin{figure}
\includegraphics[width=\columnwidth]{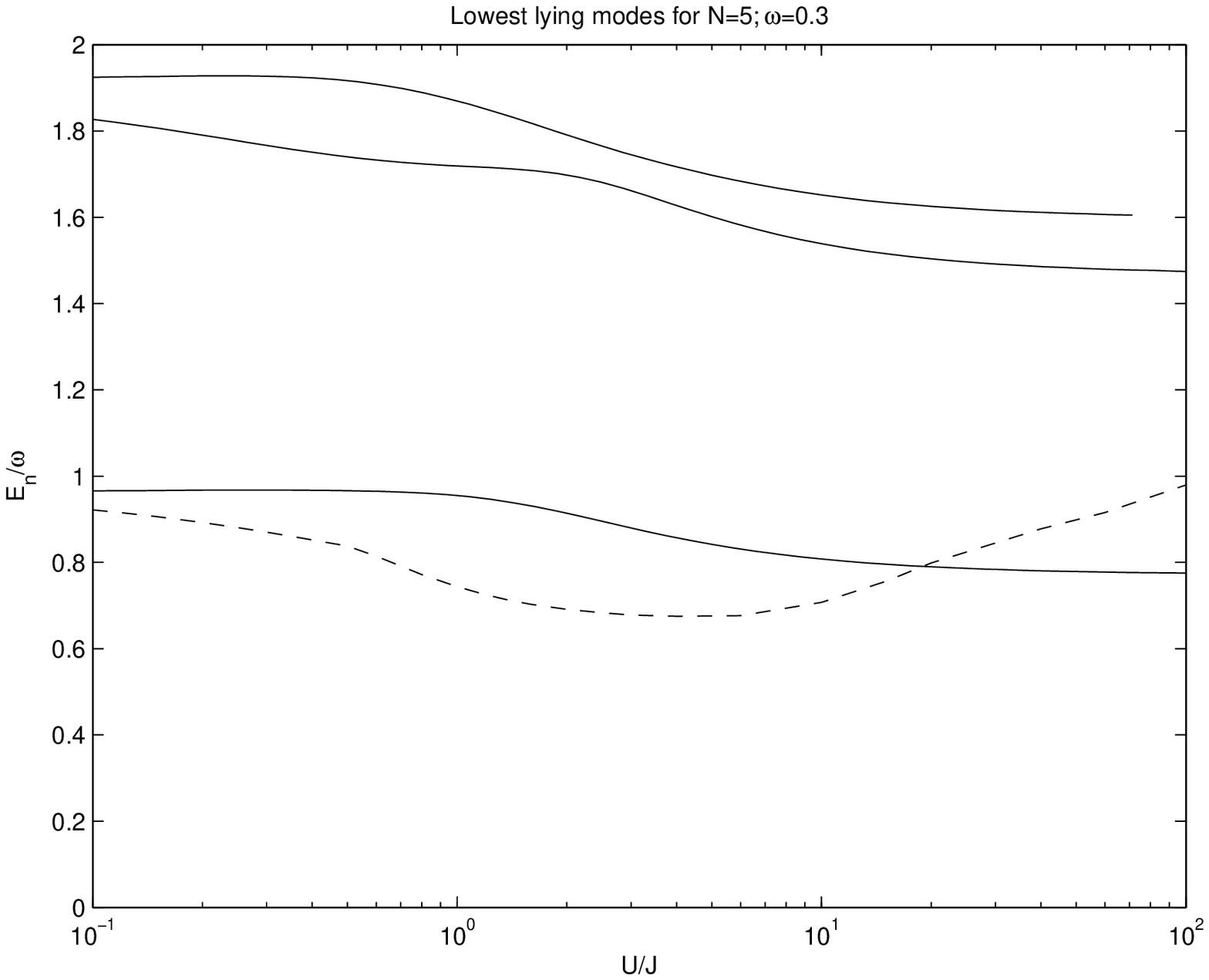}
\caption[]{Lowest excitation frequencies in units of the trap
frequency $\omega$ in a shallow trap with $\omega=0.3$. Full
lines represent the results for the three lowest excited states
obtained by exact diagonalization.
Dashed line is the
outcome of the Bogoliubov approximation.
\label{fig:modesshallow}}
\end{figure}

\begin{figure}
\includegraphics[width=\columnwidth]{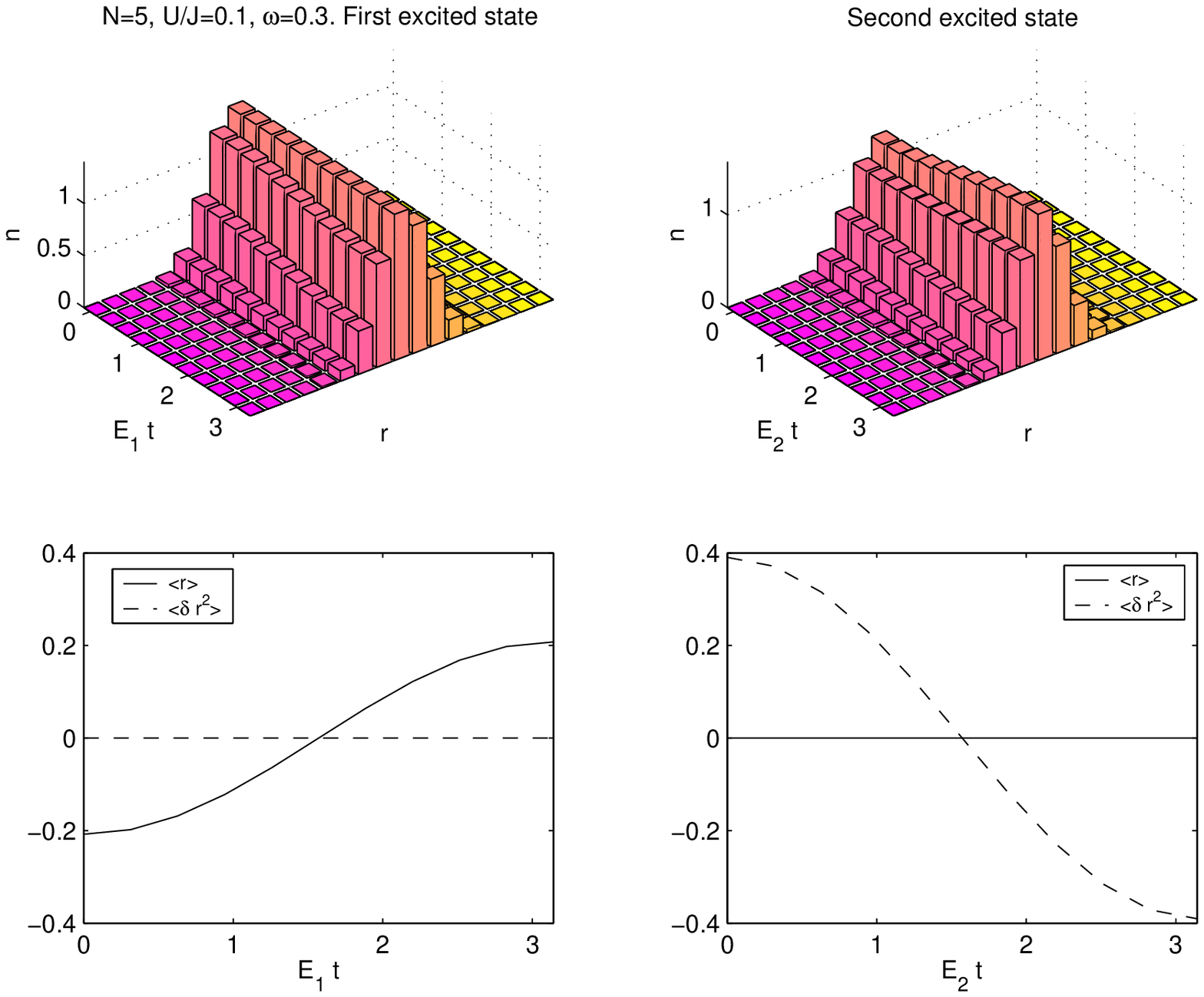}
\caption[]{Time dependence of the density profile for a system
with $U=0.1$ and $\omega=0.3$ (weak trapping and weak interactions).
The different sets of bars show the density at successive time
instances for a system prepared in a superposition of the ground
state and an excited state (left panel, first excited state;
right panel, second excited state) with the amplitudes 0.92 and
0.4, respectively. The curves in the lower two
panels show the variation of the center-of-mass position and
mean squared radius with time, showing the dipole and breathing
character, respectively, of the oscillations. To enhance
visibility, the curve for the mean squared radius has been shifted
down by an amount equal to its time
average, defining the plotted quantity as
$\langle \delta r^2\rangle = \langle\psi(t)|r^2|\psi(t)\rangle-
\bar{r^2}$ where $\bar{r^2}$ is the time average.
\label{fig:oscshallow0.1}}
\end{figure}

The mode frequencies, i.~e.\ the excitation energies relative
to the ground state, for the
lowest four excited states are displayed in Fig.\
\ref{fig:modesshallow}.
When the coupling becomes weak, the mode frequencies approach
integer multiples of the trap frequency with the expected
degeneracies in a harmonic trap. Restoring dimensions, with the
aid of Eqs.\ (\ref{jdef},\ref{omegadef}), we find in fact
\cite{cataliotti,kraemer2002a}
\beq
E_n = q \sqrt{\frac{m}{m^*}}\hbar \Omega,
\enq
with $q$ integer. 
Clearly the the integer level spacing is approximate, as is
the degeneracy, because in the $N=5$ case discreteness
is still manifest.
The anticipated dipole
and monopole (breathing) oscillations are visualized in Fig.\
\ref{fig:oscshallow0.1}, where the density evolution in time of a
superposition of the ground state and each excited state is shown
together with the mean position and width of the cloud.

\begin{figure}
\includegraphics[width=\columnwidth]{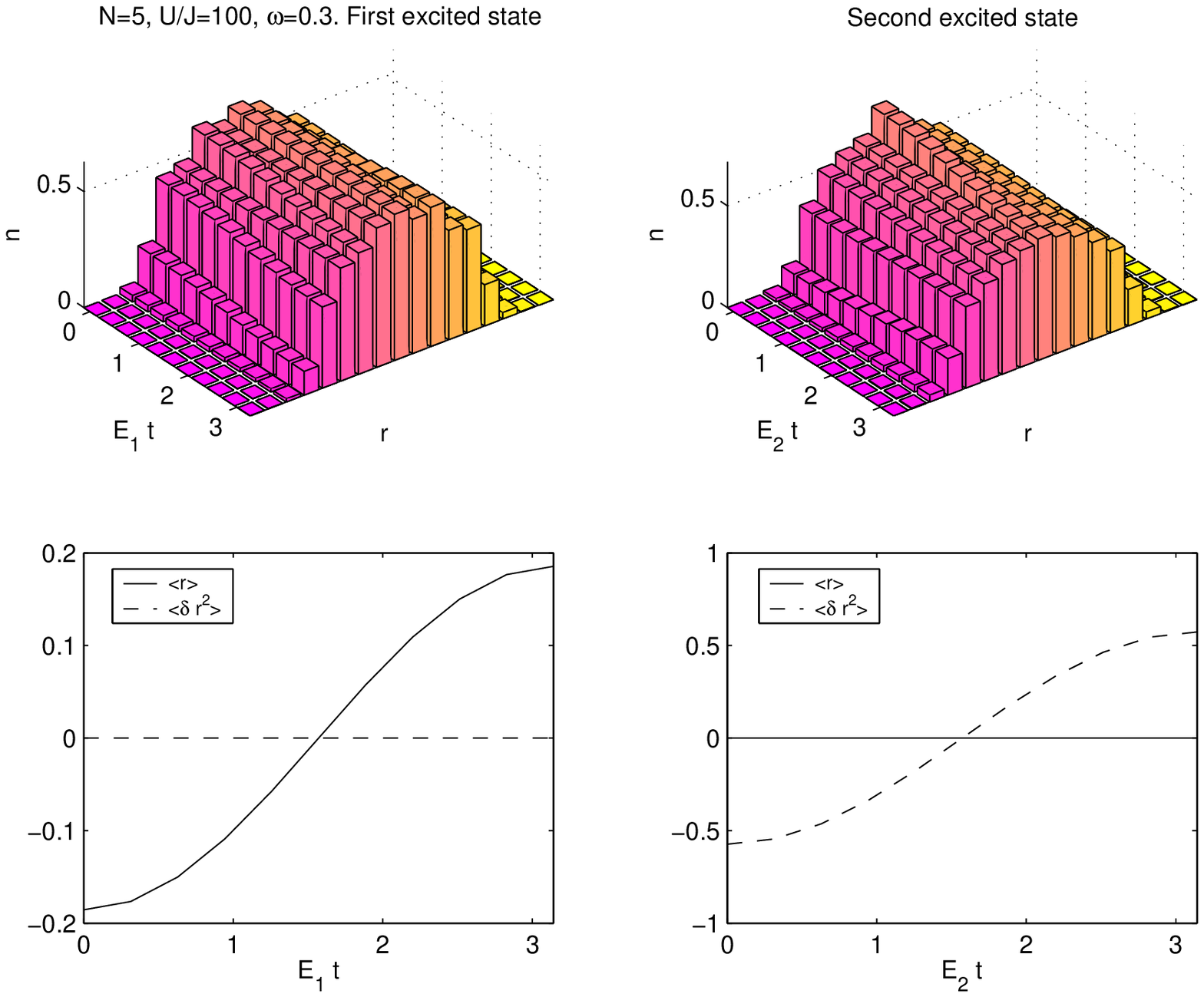}
\caption[]{Same as Fig.\ \ref{fig:oscshallow0.1}, but here
the coupling is $U=100$.
\label{fig:oscshallow100}}
\end{figure}

In a trapped BEC, as the coupling grows stronger the frequency of
the dipole mode stays constant while that of the monopole mode
decreases \cite{pethick2001a}; in the Thomas-Fermi limit in one
dimension it is down to $E_2=\sqrt{3}\omega \approx 1.73\omega$. In
the present system, one cannot hope to see this limit, but there
is indeed a drop of the second excitation frequency while
the lowest one initially stays constant. As the coupling strength approaches
and exceeds unity, however, the discreteness of the system becomes
manifest in a $U$-dependence, although weak, of the dipole
frequency. The dipole and monopole character of the first and
second excited state, respectively, are unchanged, as is exemplified 
in Fig.\ \ref{fig:oscshallow100}.

In the limit of weak interactions, one expects the Bogoliubov
approximation to be valid. This approximation can for an
inhomogeneous system be effected by
linearizing the Gross-Pitaevskii equation around its ground-state
solution \cite{pethick2001a}. The latter equation is obtained
by replacing the field operators in the Hamiltonian, Eq.\
(\ref{hamiltonian}), by their expectation values,
$a_r \to z_r=\langle a_r\rangle$, resulting in the discrete
difference equation
\beq
U|z_r|^2z_r -\frac{J}{2} (z_{r+1}+z_{r-1}) +
\frac{\omega^2}{2}r^2 z_r.
\enq
In fact, solving the full Gross-Pitaevskii equation also for the 
excited states turned out to be easier than
diagonalizing the linearized Bogoliubov equations, and therefore the
former method was chosen. The result thus obtained for the 
first excited state is included in
Fig.\ \ref{fig:modesshallow}, and it can be seen that the Bogoliubov
approximation fails appreciably even for weak interactions.
The reason is that the derivation of the Gross-Pitaevskii
equation assumes that the number of atoms in the ground state,
$N_0$, greatly exceeds the quantum fluctuations around it,
which, however,
are of order unity; for the case of five particles this condition
is certainly not met.\footnote{It is important here to make a
distinction between quantum fluctuations in different quantities.
In a Bose-Einstein condensed state, the occupation of the lowest
single-particle state is macroscopic, and the quantum fluctuations
around this solution are small (hence it is in field theory
called a classical solution). On the other hand, the quantum
fluctuations in the occupation of each site are big in this regime,
unlike in the Mott state.}

\section{Modes in tight traps}
\label{sec:tight}

\begin{figure}
\includegraphics[width=\columnwidth]{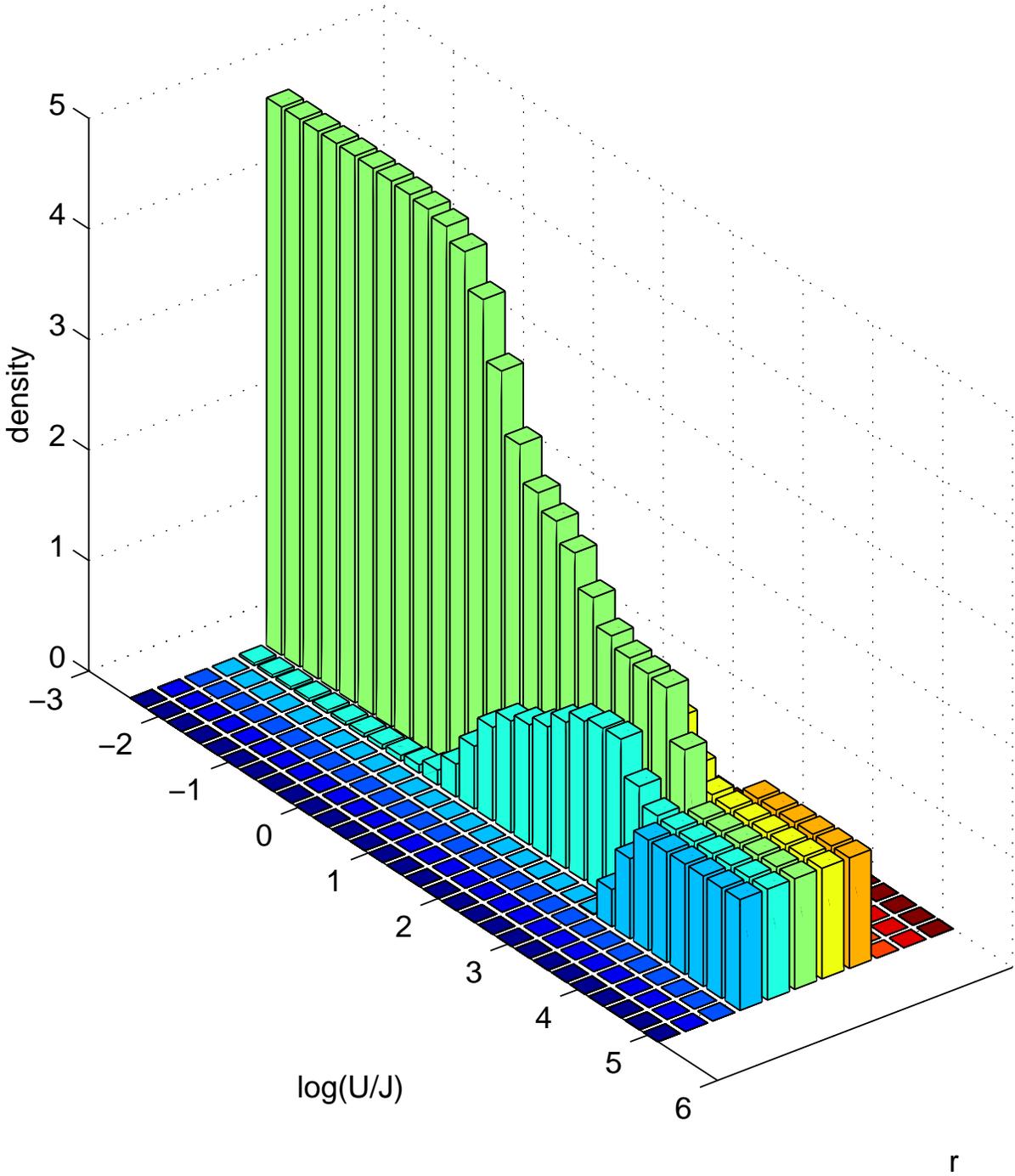}
\caption[]{Ground-state density in the case of a tight trap,
$\omega=4.0$.
\label{fig:gdenstight}}
\end{figure}

Let us now turn to the case of a tight trap. Figure \ref{fig:gdenstight}
displays the ground-state density distribution for values of $U$
ranging from weak to strong interactions, in a trap with frequency
$\omega=4.0$.
The physics is here determined by the balance between interactions
and trap and the tunneling has little effect.
The phase diagram has much more structure in this limit compared
to the shallow-trap case.

\begin{figure}
\includegraphics[width=\columnwidth]{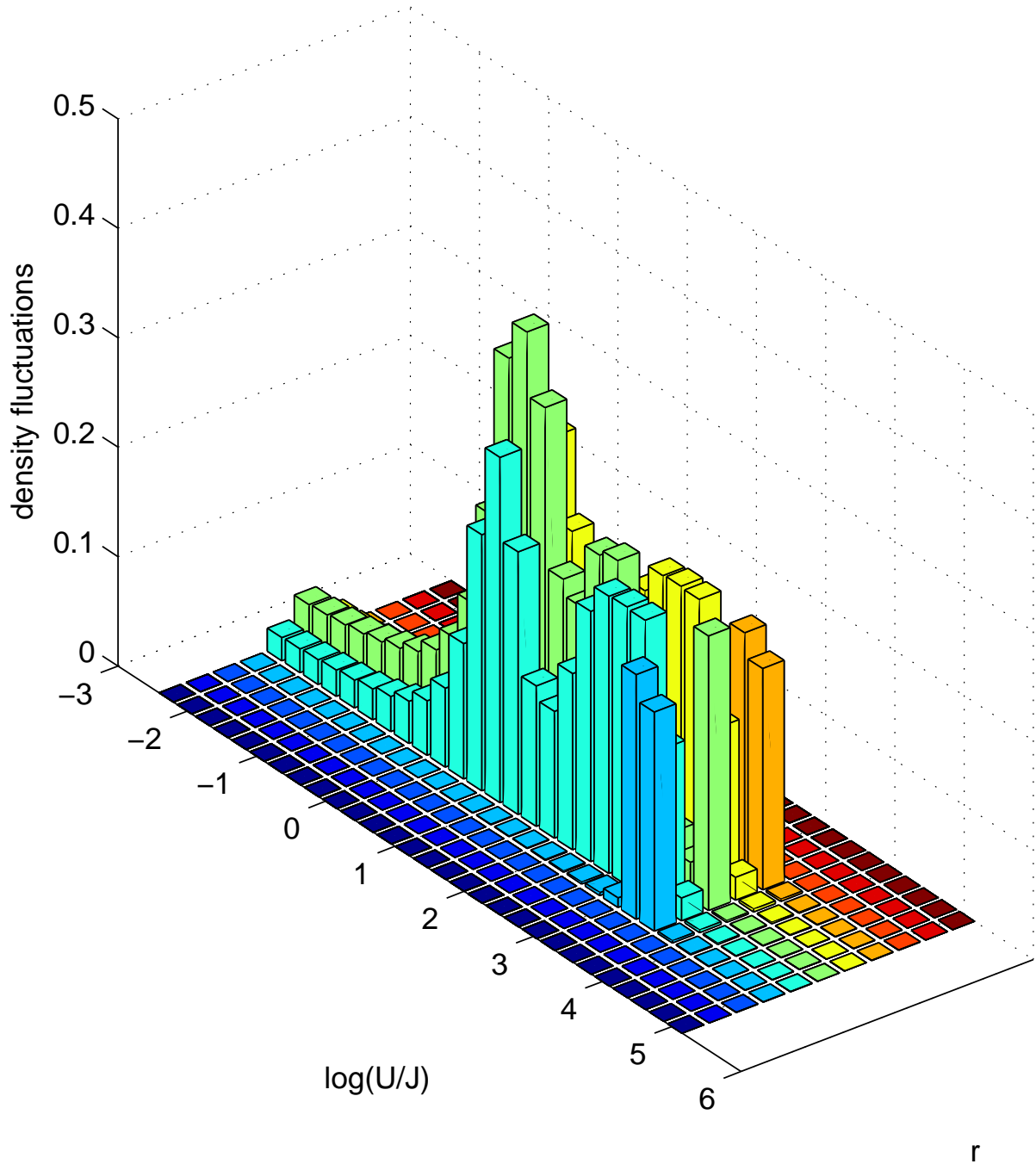}
\caption[]{Quantum fluctuations in the ground-state density
in a tight trap, $\omega=4.0$.
\label{fig:flucttight}}
\end{figure}

For weak
interactions ($U \lesssim 0.1$), all the particles simply gather
in the central well (or the two central wells, if $L$ were even).
With increasing $U$ the density distribution flattens, and
for strong enough interactions the ground state resembles
a Mott insulating state, with one particle in each of the centermost
wells. (If the even/odd parity of the number of wells and the 
number of particles 
do not match, the edge sites will be partially filled.)
Figure \ref{fig:flucttight} verifies
that in this state the quantum fluctuations are minimal, which
warrants the use of the term Mott state.
The transition to the Mott state in this five-particle system
is in fact quite sharp. Also in the limit of
small $U$, the quantum fluctuations become small, but they only
vanish completely in the limit $\omega\to\infty$ and the transition
is not sharp.

Between the two extremes, the shape of the atom cloud is
determined by a balance between interactions and trap potential.
It is quite easy to estimate the crossover values of $U$ where
the system changes between different types of ground state (we
avoid speaking of phases for this finite system).
If the
effective size of the system is $R$ sites, the interaction energy
$E_i$ and trap energy $V$ scale as
\beq
E_i\sim \frac{N^2U}{R} ,\, V\sim \omega^2R^2.
\enq
Balancing these yields $R^3\sim N^2U/\omega^2$ or
\beq
U=\frac{\omega^2 R^3}{N^2}.
\enq
The Mott state, where $R=N$, thus sets in above
$U=\omega^2 N$; inserting the present parameters we get $U=80$.
When the particles gather in the centermost well, $R$ is equal to 1;
this happens when
$U=\omega^2/N^2\approx 0.5$ or smaller. This back-of-the-napkin
argument is in almost quantitative agreement with the exact numerical
findings.

\begin{figure}
\includegraphics[width=\columnwidth]{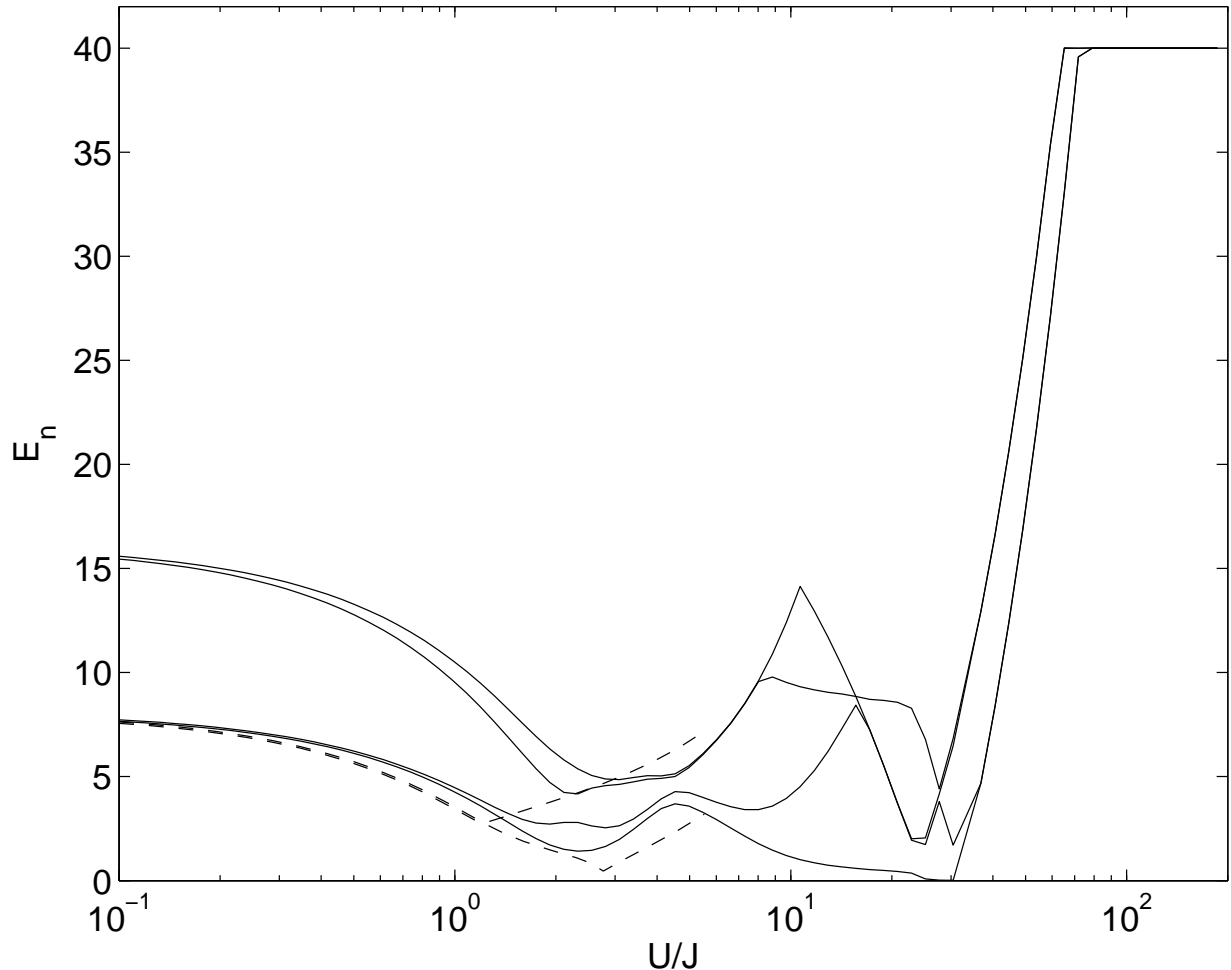}
\caption[]{Lowest few excitation frequencies in a tight trap,
$\omega=4.0$. Dashed lines represent the Bogoliubov approximation.
\label{fig:modestight}}
\end{figure}

\begin{figure}
\includegraphics[width=\columnwidth]{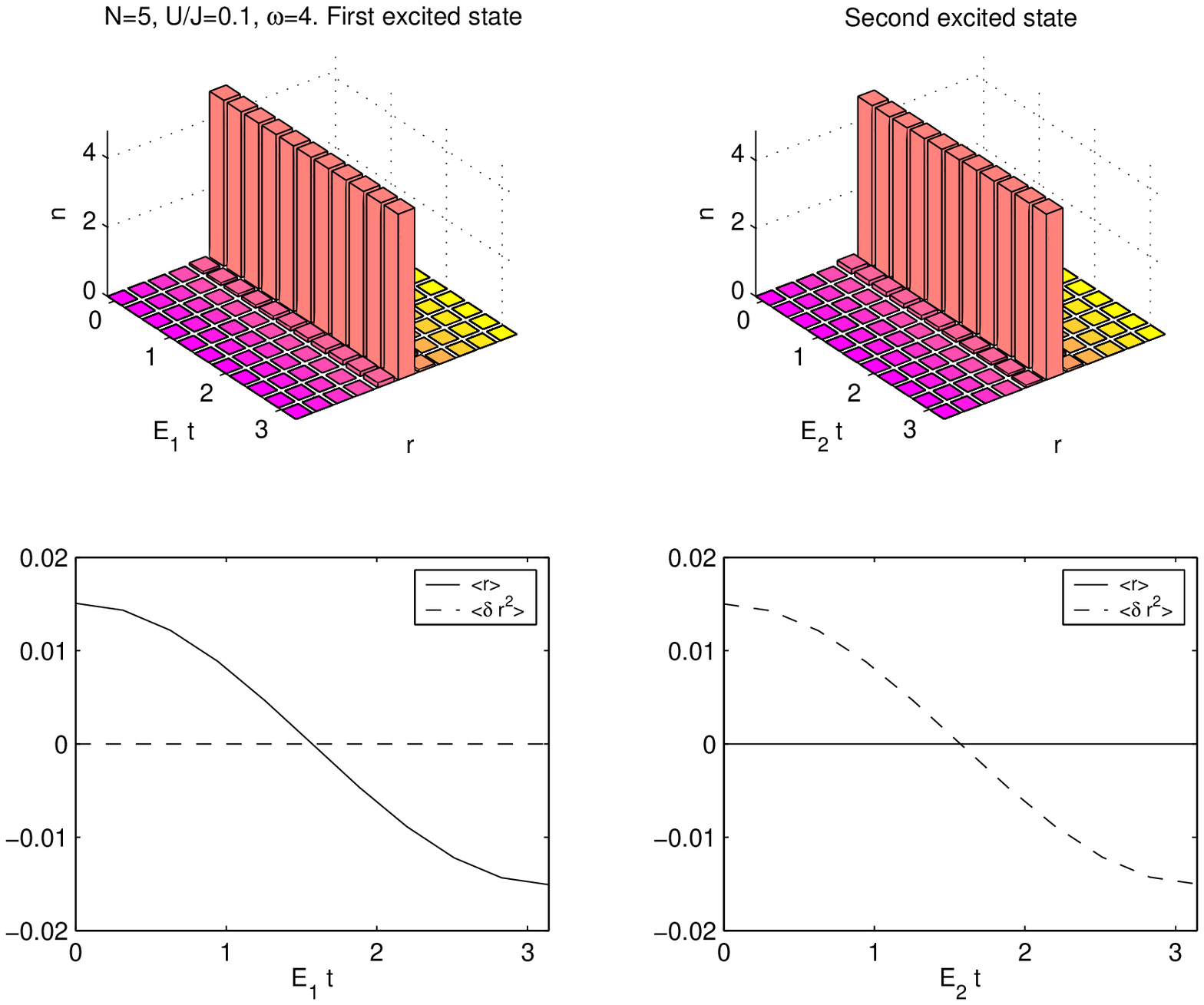}
\caption[]{Visualization of the two lowest modes for a tight trap,
$\omega=4.0$, and weak coupling, $U=0.1$. Panels are as
in Fig.\ \ref{fig:oscshallow0.1}.
\label{fig:osctight0.1}}
\end{figure}

The lowest few mode frequencies as
functions of $U$ when $\omega=4.0$ are displayed in Fig.\
\ref{fig:modestight}. (Observe that, in order to emphasize
physical interpretation, the frequencies were in 
Fig.\ \ref{fig:modesshallow} given in units of the trap 
frequency $\omega$, but here it is given in units of 
the tunneling $J$).
The dependence on coupling seen in
Figs.\ \ref{fig:gdenstight}-\ref{fig:flucttight} is seen to have clear
consequences also for the mode frequencies.
In the central-well limit,
$U\lesssim 1$, the frequencies are easily interpreted since the
trap determines all the physics, and all the
eigenstates of the Hamiltonian are pure Fock states or
superpositions of degenerate Fock states.
The lowest two modes are superpositions
of the two possible states that result when one particle
is removed from
the central site and put in one of the two neighboring ones,
$|\ldots 00410\ldots\rangle$ and $|\ldots 01400\ldots\rangle$.
The limiting value of the excitation frequency
is equal to $\omega^2/2=8$, the
excess energy of one particle being moved to a neighboring site.
The energy of the degenerate third and fourth excited states is
twice this, since they have
two particles in in an off-center well.
One may expect that among the two lowest excitations,
the superposition with a minus sign
corresponds to a dipole mode and the plus sign corresponds to a
breathing mode. Indeed, this is confirmed in Fig.\
\ref{fig:osctight0.1}, where the time evolutions of the modes are
visualized, but note the small amplitude of the oscillations.
The time dependence is merely due to the
small deviations from perfect confinement that are still left in
the moderately strong trap with frequency $\omega=4.0$; it can
easily be realized that no current can flow in a system whose
eigenstates are pure Fock states in configuration space. To see
this, consider preparing
an initial state as a superposition of two eigenstates of the
system, $|\psi\rangle=\alpha|A\rangle+\beta|B\rangle$, and study
its time evolution,
\beq
|\psi(t)\rangle = e^{-iE_At}\left(\alpha |A\rangle +
e^{-i\omega_{AB}t}\beta|B\rangle\right),
\enq
where $\omega_{AB}=E_B-E_A$.
Now expand the time evolution in Fock states $|f\rangle$, and
obtain for the density at the position $r$,
\beq
\langle \psi(t) |\ad_r a_r| \psi(t) \rangle = \sum_f
|\alpha\langle f | A \rangle+e^{-i\omega_{AB}t}\beta\langle f | B \rangle|^2
\langle f |\ad_r a_r|f\rangle,
\enq
and we see immediately that if $|A\rangle$ and $|B\rangle$ have no
Fock state components in common, there can be no time dependence.
This fact lies behind the insulating nature of
the Mott state, but as we have seen, it also prohibits the
dynamics in the limit of a very strong trap.

\begin{figure}
\includegraphics[width=\columnwidth]{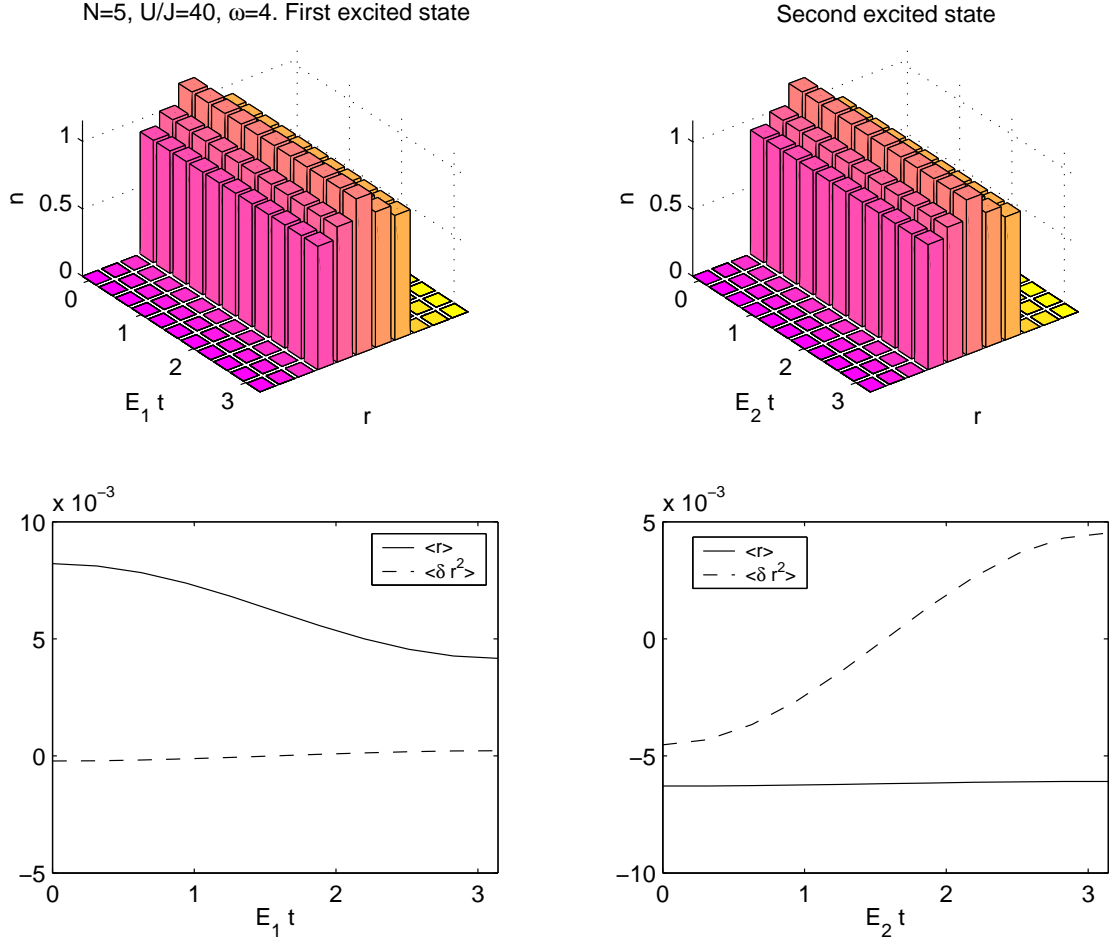}
\caption[]{Same as Fig.\ \ref{fig:oscshallow0.1}, but here
the trapping frequency is $\omega=4.0$ and the coupling is $U=40$.
\label{fig:osctight40}}
\end{figure}

As the coupling gets stronger, the degeneracy is lifted and an
intricate pattern of level crossings follows; however, the
dipole/monopole character of the two lowest modes is retained
until $U$ approaches the value of approximately 50, where a mixing
of the dipole and monopole modes starts to be visible, as seen in
Fig.\ \ref{fig:osctight40}.
When $U \gtrsim 100$, we are in the Mott state and the mixing is
complete so that the four lowest-lying modes are degenerate
with an energy equal to $5\omega^2/2$; this is the energy
difference between the second and third well from the center as
each excited state corresponds just to a displacement of a
particle from the center.
\begin{figure}
\includegraphics[width=\columnwidth]{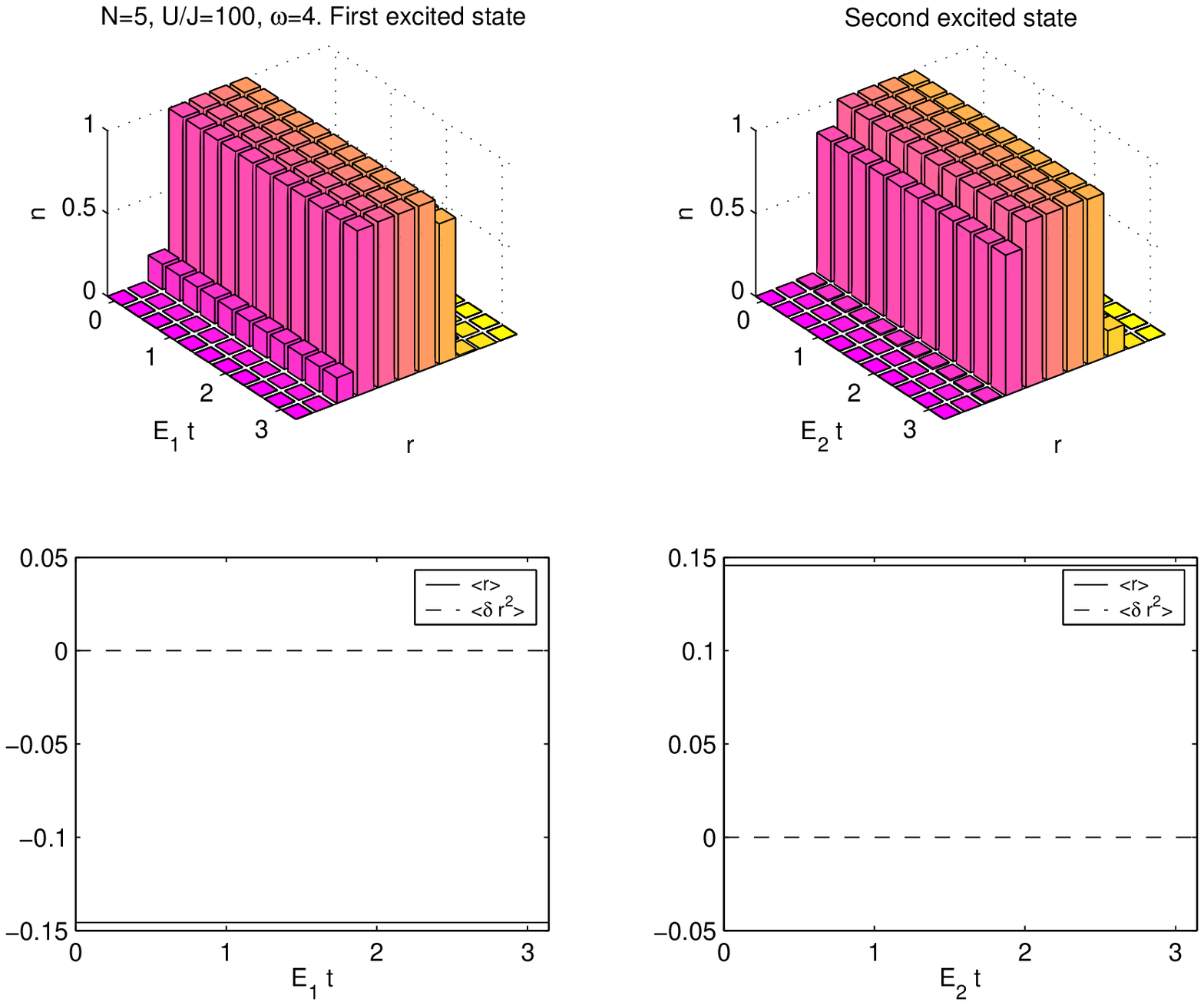}
\caption[]{Same as Fig.\ \ref{fig:oscshallow0.1}, but here
the trapping frequency is $\omega=4.0$ and the coupling is $U=100$.
\label{fig:osctight100}}
\end{figure}
In fact, each of the two lowest
excited states has very high overlap (the overlap is
0.98 when $U=100$) with one
of the pure Fock states $|\ldots 011111000 \ldots\rangle$ and
$|\ldots 000111110 \ldots\rangle$, i.~e., the ground-state
configuration displaced from the center by one lattice site. The
third and fourth excited states are separated from the first two
by an exceedingly small energy gap (about 0.005 in units of $J$),
and are mainly superpositions of the state
$|\ldots 010111100 \ldots\rangle$ and its mirror reflection.
As seen in Fig.\ \ref{fig:osctight100},
excitation of one of these modes does not result in oscillation;
this is again a consequence of the fact that the eigenstates are
pure Fock states in the spatial representation.
Indeed, this is a Mott insulating state and excitation does not
result in particle flow.

Returning to the mode frequency plot in Fig.\
\ref{fig:modestight}, the frequencies have also been calculated in the
Bogoliubov approximation and are included as dashed lines.
Somewhat surprisingly, the Bogoliubov approximation performs better
in the strong-trapping regime than in the weak-trapping regime,
although one would na{\"\i}vely expect the accuracy to be worse
when tunneling is suppressed: the quantum fluctuations in local 
density are
small. The apparent paradox is resolved by noting that in this limit
the system is in fact a Bose-Einstein condensate: all
particles occupy the same state, namely the one confined to the
central well.

\section{Conclusions}
\label{sec:conclusions}
The mode frequencies of a trapped boson system in an optical lattice
have been studied with attention to the dependencies on trap
strength and interactions. In the weakly-trapped limit, it is
shown that the mode frequencies have the usual character of dipole
oscillations for the lowest-lying mode and breathing oscillations
for the next-lowest, and the frequencies are, despite the low
number of particles and the discreteness, seen to approximately
approach the result for a harmonically trapped gas when the
interactions vanish. Discreteness effects set in when the
interaction energy is comparable to the tunneling. 
For sufficiently
strong trapping, the number fluctuations are quenched, and the
dynamics therefore absent, in two limits: when the
interactions are strong, even the five-particle system displays a
quite sharp transition to a Mott insulating state. When on the other
hand the trapping potential dominates completely, all the particles
are trapped in the central well and the amplitude of any oscillation
goes to zero. 
It turns out that the Bogoliubov approximation is capable
of approximating the mode frequencies better for strong trapping than 
for weak trapping: in the weak-trapping case
the number of particles is too small for the Bogoliubov approximation
to work, but a strong trap quenches the fluctuations around the 
Bose-Einstein condensed ground state.

\begin{acknowledgments}
This project was financially supported by the G{\"o}ran Gustafsson 
foundation. The author is grateful to Patrik Henelius for 
discussions.
\end{acknowledgments}


\end{document}